\documentclass[aps,prb,twocolumn]{revtex4}
\usepackage{epsfig}
\usepackage{rotating}
\usepackage{graphicx}
\usepackage{amsmath}
\newcommand \bea {\begin{eqnarray} }
\newcommand \eea {\end{eqnarray}}

\begin{document}

\title{Small polarons and c-axis transport in highly anisotropic metals}

\author{A. F. Ho$^*$\cite{currentaddress} and A. J. Schofield$^{\dagger}$}

\affiliation{$^*$ The Abdus Salam ICTP, Strada Costiera 11, 
34100 Trieste, Italy.\\
$^{\dagger}$School of Physics and Astronomy, Birmingham University, Edgbaston, 
Birmingham B15 2TT, U.K.}



\vskip 0.5 truein

\begin{abstract}

Motivated by the anomalous $c$-axis transport properties of the
quasi two-dimensional metal, $\rm Sr_2 Ru O_4$,
 and related compounds, we have studied the interlayer
hopping of single electrons that are coupled strongly to $c$-axis 
bosons. We find a $c$-axis resistivity that reflects
the in-plane electronic scattering in the low  and very high temperature
limits (relative to the characteristic temperature of the boson
$T_{\rm boson}$). For temperatures near
the $T_{\rm boson}$, a broad maximum in the resistivity can appear for
sufficiently strong electron-boson coupling. This feature may
account for the observed ``metallic to non-metallic crossover'' seen
in these layered oxides, where the boson may be a phonon.  

\end{abstract}

\maketitle



\section{Introduction}

Since the discovery of the cuprates, 
many studies have focused on the putative two dimensional (2d)
non Fermi liquid state of the copper oxide layer in the normal
state of these materials.\cite{NFL} However, the superconducting state
is three dimensional and, at least near to the superconducting critical
temperature ($T_c$),  some aspects of the physics of coupling 
between the planes must become important.\cite{Leggett} 

The transport properties in the weakly-coupled dimension (denoted  the
$c$-axis here) are also interesting in their own right\cite{CooperGray}. 
In the normal state of many of these layered copper oxides, 
the resistivity ratio,
$\rho_{ab}/\rho_c$, is of order $10^3$ to $10^5$  but intriguingly, this
ratio is temperature dependent:  $\rho_c$ is non-metallic
 near $T_c$, $\quad\rho_c \sim T^{-\gamma}, \quad 0 < \gamma <2 $, while
the in-plane resistivity has the marginal Fermi liquid behavior
$\rho_{ab} \sim T$ (at least at optimal doping)\cite{CooperGray}.  

\begin{figure}[h]
\includegraphics[width=\columnwidth]{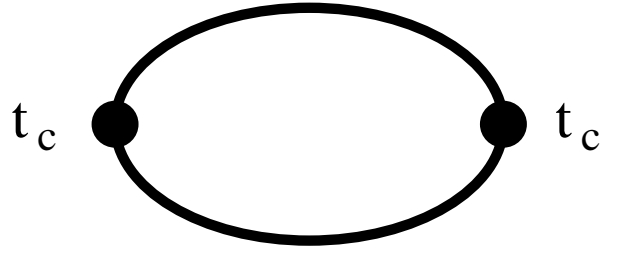}
\caption{The order $t_c^2$ contribution to the $c$-axis conductivity. The
thick lines represent the renormalised in-plane Green's function. }
\label{fig1}
\end{figure}

This dichotomous behavior  cannot easily be accommodated 
within an anisotropic three dimensional (3d) 
Fermi surface picture. We begin with the assumption that the $c$-axis
coupling is simply interlayer hopping of single electron:
\bea 
H_c = t_c \sum_n c^{\dagger}_{n+1} c_n + h.c. \label{simplehop} \; ,
\eea where $n$ is the layer index. Then, at temperatures larger than
the $c$-axis electron hopping matrix element $t_c$, simple
perturbation theory (to lowest order in $t_c$) with the Kubo formula 
gives the $c$-axis
conductivity to be proportional to the polarization bubble (Fig.~\ref{fig1}).
In the figure, the thick lines represent the renormalised in-plane Green's
function. Hence the $c$-axis
conductivity directly reflects the {\it in-plane} spectral weight.
But if the in-plane state is a metal, then the $c$-axis
conductivity should also be metallic.

There are at least two ways to bypass this argument. Either
the in-plane physics is unusual or the interlayer coupling
is unconventional. Considering the first possibility, there
might be  a non Fermi liquid in-plane Green's function\cite{PWAbook}.
More conventionally, 
Ioffe and Millis\cite{Ioffe} argued for superconducting
phase fluctuations, and others have just assumed
some phenomenological form of in-plane spectral weight which has
strong in-plane scattering built-in\cite{Littlewood}.  In
such schemes, the interlayer transport directly probes the anomalous
in-plane scattering.
Within the second possibility, Rojo and Levin\cite{Levin}
considered static disorder, and also boson-assisted hopping in the
$c$-axis direction,  while Turlakov and Leggett\cite{Turlakov} studied
interplane and in-plane charge fluctuations. These approaches
 can lead to incoherent $c$-axis transport,
without the need to invoke  strongly anomalous in-plane behavior.


Motivated by experiments in a related system,
the strontium ruthenate family,
we follow the second route in this paper and
 study  a boson-assisted hopping mechanism for the unusual 
$c$-axis transport in the ruthenate systems. 

In contrast to previous work on boson-assisted 
hopping\cite{Levin},
in this report we treat the  strong  $c$-axis electron-boson coupling 
{\it exactly} via a canonical transformation,  
 and then add the weak $c$-axis hopping perturbatively. 
Thus the essential physics is that of the small
polaron.\cite{Holstein,foot} 
 This can be seen most easily after the canonical
transformation: when the electron hops from one plane $n+1$ to
another at  $n$, a cloud of bosons are created or destroyed:
\bea
c^{\dagger}_{n} c_{n+1} 
\overset{\mathrm{canon.}}{\underset{\mathrm{transf.}}{\longrightarrow}}
 c^{\dagger}_{n} c_{n+1}
\exp \left[ \sum_{q} A_n(q) ( a_q - a^{\dagger}_{-q} ) \right],
\eea (The precise form of the amplitude $A_n(q)$ is shown in
Eq.~\ref{H_can} below.) At temperatures lower than the characteristic
energy of the boson, $T_{\rm boson}$, 
electrons can hop from one plane to another, but this hopping
is suppressed due to the ``dragging'' of the boson cloud. With increasing 
$T$,  more and more bosons are created and destroyed; due to this inelastic
scattering process, the hopping electron acquires a large imaginary part
in the self-energy: interlayer hopping becomes more like diffusive.
Hence a crossover into a non-metallic behaviour occurs. 

Note that there is a crucial separation of scales. We assume 
that, because of  the
highly anisotropic electronic dispersion in these quasi-$2d$ materials,
$\epsilon_{\rm Fermi}^{\rm in-plane} \gg T_{\rm boson} \gg t_c$. Thus, it is 
possible for the strong electron-boson coupling to 
show  up dramatically in $c$-axis
transport, while the in-plane behavior is relatively unaffected. Also,
it allows us to treat the $c$-axis hopping perturbatively and still
have quantitative control over the crossover region near $T_{\rm boson}$. 

In this paper, we shall concentrate on the d.c. resistivity in the 
$c$-direction. Our main finding is, indeed, that due to the 
small polaron, a metal to non-metal crossover 
for temperatures near the boson scale $T_{\rm boson}$ is obtained.
 At higher or lower
temperatures, the $c$-axis transport basically reflects the in-plane
dynamics through the in-plane scattering rate.

We have so far  not specified the nature of the boson. This is because
 the analysis to be presented can, in general, be applied
 to any kind of strong electron-boson coupling, 
provided there is this separation of scales mentioned,
and the bosonic mode is gapped and neutral. 
For the rest of the paper, we shall phrase
the analysis in terms of  phonons, with a view towards application to 
some members of the strontium ruthenate family. For other members of the same
family with known magnetic instabilities, and perhaps for the
cuprates, this boson could be a paramagnon or other magnetic excitation.  

Strontium ruthenate $\rm Sr_2 Ru O_4$
is iso-structural to one of the parent compounds of the cuprates, 
$\rm La_2 Cu O_4$. Above its
superconducting transition temperature of $1.5 K$, it is a well-characterised
3d Fermi liquid\cite{dHvA}, and yet, above a coherence scale of $\sim 25 K$,
 the $c$-axis optical conductivity loses its Drude peak\cite{optics}, and
above a ``cross-over'' scale, $T_{\rm max} \sim 130 K$, the $c$-axis resistivity
becomes non-metallic while the $ab$-plane resistivity stays metallic,
albeit with a  close to linear temperature dependence and a value 
which eventually exceeds the Mott-Ioffe-Regel limit\cite{resist}.  
In the recent report of Jin {\it et. al.}\cite{Jin} in the related system 
$\rm Ca_{1.7} Sr_{0.3} Ru O_4$, where there is the same dichotomous resistivity
behaviour, $T_{\rm max} \sim 190 K$ appears to be tied to a structural 
phase transition. In $\rm Sr_2 Ru O_4$, there is no such direct link, but 
there are strong electron-phonon interactions in the 
$c$-direction.\cite{Braden} This motivates us to consider
phonon-assisted  hopping as a potential origin for the 
non-metallic $\rho_c$ in these systems
at high temperatures.

A further experimental observation in the vicinity of $T_{\rm max}$ in
the $c$-axis resistivity is that the magnetoresistance can change sign
and becomes negative above $T_{\rm max}$. This happens\cite{Hussey} in both the
longitudinal (field parallel to the current) and transverse 
(field perpendicular to the current) directions in $\rm Sr_2 Ru O_4$. 
Somewhat similar properties can be found in some of the cuprates,
although there seems to be no universality in the $c$-axis 
magnetoresistance between the various families of the 
cuprates.\cite{cupratesMR} In this paper, we shall calculate
the orbital component of the magnetoresistance. This will necessarily
be positive and will not explain these observations. However, a more
complete theory would need to take into account the effect of magnetic
field on the propagators and the bosonic mode, which is beyond the
scope of this paper, but we will argue that it must be these
contributions that change the sign of the magnetoresistance, 
within this picture at least.
 


The plan of the paper is as follows: in Section II, we define our model,
in Section III, we study the model in the strong electron-phonon coupling
limit. Section IV gives the results.
Section V discusses the relevance
of this model to the strontium ruthenate family and beyond.

\section{Model}

As already emphasized in the
Introduction, the analysis to be presented can be applied generally,
with little modification,
to any strong coupling of electrons to a gapped, neutral bosonic mode, but
to be concrete, we shall consider coupling to phonons. 

Our model involves electrons coupled to the polarisation induced by the out of
phase vibration of the ions within a unit cell; the form of this electron-optical
phonon coupling has been derived long ago in the $3d$ case\cite{frohlich}:
\bea
H^{\rm 3d}_{\rm e-ph} = \sum_n \int d^2 x \sum_{\vec q}  
M_{\vec q} \exp(i {\vec q} {\vec R}_n) \rho_n({\vec x})
 \left(a_{\vec q} + a^{\dagger}_{-{\vec q}}\right)  ,
\eea where  ${\vec R}_n = [{\vec x}, z_n]^{T}$, ${\vec x}$ is the intraplane
coordinate, $z_n  = n c$ is the plane coordinate, with $c$ the 
interlayer distance and $n$ the plane index. $\rho_n({\vec x}) = 
c^{\dagger}_{n}(\vec{x}) c_{n}(\vec{x})$ is the electronic density. 
The electrons mainly couple to longitudinal phonons (represented by 
$a_{\vec q} , a^{\dagger}_{\vec q}$) because only these
phonons set up strong enough electric fields when they vibrate\cite{Mahan}. Then the
direction of the polarisation vector locally is simply along the direction given by the
vector sum of the displacements of the ions in a unit cell.
Now, for the mode that can couple to $c$-axis charge transport, 
we expect that in these highly anisotropic metals such as the cuprates and
$\rm Sr_2 Ru O_4$,  the effective force constant between the planes may be
less stiff than those  within the plane. This is because of the strong
covalent bonding and the strong band electronic correlation within the plane, as compared
to the weaker bonding and correlations between the planes. 
Effectively,  due to the large in-plane 
effective force constants, we let the vibration frequency for
the in-plane modes $\omega_{{\vec q}} \longrightarrow \infty$ compared to the
interplanar one. In any case, we wish to separate out the direct effect of strong 
electron-phonon coupling on $c$-axis transport, from the effect of phonons on 
in-plane propagation which then influences the $c$-axis transport via the mechanism
sketched in Fig.~\ref{fig1}.   
Thus, with such a simplification, the Fourier transform of the displacement 
vector ${\vec{Q}}$ is only in the interplane direction: 
$ {\vec{Q}} \propto \frac{\vec{q}}{q} (\omega_{{\vec q}})^{-1/2} 
(a_{\vec q} + a^{\dagger}_{-\vec{q}}) \approx [0,0,Q_c]^{T}$, with
$Q_c \propto (\omega_{ q})^{-1/2}  (a_q + a^{\dagger}_{-q})$, and the propagation
direction $q$ (not a vector!) is along the $c$-axis. 
In effect, all the ions in a plane oscillate together rigidly, in the direction
perpendicular to the plane, and the propagation direction is also parallel to the $c$-axis. 
Hence it is the total charge 
$\rho_n = \int d^2 x  c^{\dagger}_{n}(\vec{x}) c_{n}(\vec{x})$ of the $n$'th plane that
couples to the phonon creation and annihilation operators $a^{\dagger}_q, a_q$:
\bea
H_{\rm e-ph}  & = & \sum_n \sum_q  M_q \exp(i q z_n) \rho_n
 \left(a_q + a^{\dagger}_{-q}\right)  ,
\eea where we have combined all the prefactors as usual into the electron-phonon
coupling $M_q$.

If the bosonic mode is of magnetic origin, we may argue that because of stronger
electronic overlaps in-plane compared to the interplane direction, 
and hence stronger in-plane exchange effects, the magnetic fluctuation mode is 
softer in the interplane direction, and again
this simplifed form of electron-boson coupling will be appropriate.

Our model Hamiltonian is thus: 
\bea \label{model}
H & = & \sum_{n} H^{(n)} + H_c + 
H_{\rm e-ph} + H_{\rm ph}   ,\nonumber \\
H_c & = & \int d^2 x \sum_n t_c c^{\dagger}_{n+1}(\vec{x}) c_{n}(\vec{x}) 
+ {\rm H.c.} \quad ,\nonumber \\
H_{\rm e-ph}  & = & \sum_n \sum_q  M_q \exp(i q z_n) \rho_n
 \left(a_q + a^{\dagger}_{-q}\right),
\nonumber \\
H_{\rm ph}  & = & \sum_q \omega_q a^{\dagger}_q a_q  . 
\eea

The model system consists of a stack of $2d$ planes described by
$H^{(n)}$.
 We shall take this as a phenomenological input, characterised by
a $2d$ Green's function or equivalently, a $2d$ spectral weight. 

In $H_c$, electrons are assumed to hop from one plane directly to a 
neighbouring plane only,  at the same $\vec{x}$; in momentum space, 
this means that the $c$-axis hopping 
matrix element $t_c$ is independent of the  in-plane momentum.
 (This ignores the complication of staggered planes
in the cubic perovskite structure  and the multiband
nature of $\rm Sr_2 Ru O_4$, see Discussion.) For
simplicity, we have left out the spin index on the electron operators 
$ c_{n}(\vec{x})$.

$H_{\rm e-ph}$ describes the electron-optical phonon interaction 
in the $c$-direction, as detailed above.
We have left the electron-phonon coupling
$M_q$ unspecified: as we shall see, it enters only within 
 an electron-phonon parameter [$\Delta$ or $\gamma(T)$
depending on the phonon dispersion, see Methods].

For $H_{\rm ph}$, The $c$-axis optical phonons have a dispersion $\omega_q $;
in this paper we shall look closely at the Einstein phonon 
$\omega_q \sim \omega_0$. This is both because optical phonons tend to have little
dispersion relative to the acoutic ones, and also we are able to obtain many 
explicit analytical results. We shall also discuss somewhat more qualitatively 
the opposite case  of a generic dispersion where the
phonon density of state does not have any sharp features.  These two forms
of dispersion can be thought of as opposite limits: the Einstein
phonon is the limit where all phonon modes (of different
wavevector $q$)  have the same energy,
while the more general dispersion corresponds to the limit when
all phonon modes are non-degenerate. 


\section{Methods}

\subsection{Canonical Transformation}
 
The key physics we are looking for is the effect of strong electron-phonon
coupling  on the charge transport in the weakest direction, the 
$c$-axis. Physically, the motion of the electron is accompanied by
the emission and absorption of a large number of phonons due to the strong
coupling, forming the so-called small polaron\cite{Holstein}. Technically,
one must deal with the large term $H_{\rm e-ph} $ first, 
and add the inter-plane
tunneling $H_c $ later as a perturbation.  The former is accomplished by the 
canonical transformation, $\bar{H} = \exp(-S) H \exp(S)$,  a straightforward 
generalisation of the transformation
well-known in the small polaron problem\cite{Firsov,Mahan}:
\bea
S & = & \int d^2 x S(\vec{x}), \nonumber \\
S(\vec{x}) & = &
- \sum_{n, q} \frac{M_q}{\omega_q} e^{i q R_n} 
c^{\dagger}_{n}(\vec{x}) c_{n}(\vec{x}) \left(a_q - a^{\dagger}_{-q}\right) .
\eea
One finds then,
\bea \label{H_can}
\bar{H} & = & \sum_{n} H^{(n)} + \int d^2 x  \bar{H}_c(\vec{x})  +
H_{\rm ph} ,\\
\bar{H}_c(\vec{x}) & = &
- \sum_{n, q} \frac{|M_q|^2}{\omega_q} c^{\dagger}_{n}(\vec{x}) 
c_{n}(\vec{x}) \nonumber \\
 & + &\sum_n t_c c^{\dagger}_{n+1}(\vec{x}) c_{n}(\vec{x}) X^{\dagger}_{n+1} X_{n} 
+ {\rm H.c.} \quad ,\nonumber \\
X_n  & = & \exp \left\{ \sum_{q} \frac{M_q}{\omega_q} e^{i q R_n}
 \left(a_q - a^{\dagger}_{-q}\right) \right\}  .\nonumber 
\eea Thus strong electron-phonon coupling leads to, (1) a renormalisation
of the in-plane chemical potential (first term of $\bar{H}_c$), which
we shall henceforth ignore, and (2) an effective vertex correction for
the $c$-axis hopping $t_c$ (second term of $\bar{H}_c$). Note that since the phonon modes
couple to the {\it total} charge density in a plane, 
$\bar{H}^{(n)} = \exp(-S) H^{(n)} \exp(S) =  H^{(n)}$: the in-plane motion of the electron
is not in the presence of the interplane polaronic effects, within this simplifed model.

\subsection{Linear Response: conductivity and magneto-conductivity} 

In this paper, we look at the zero-frequency conductivity and the
magnetoconductivity. For the conductivity, since the charge in the $n'$th
plane is $Q^{c}_n = e \int d^2 x c^{\dagger}_{n}(\vec{x}) c_{n}(\vec{x})$,
the current in the $c$-direction is just $j^{c}_n = \partial_t Q^{c}_n
= -i [  Q^{c}_n , H ]$. After the canonical transformation,  
$\bar{j}^{c}_n = -i [  Q^{c}_n , \bar{H}_c ]$, and so:
\bea
\left\langle \bar{j}^{c}_n \right\rangle = i e t_c \int d^2 x 
\left\langle \left[
c^{\dagger}_{n+1}(\vec{x}) c_{n}(\vec{x}) X^{\dagger}_{n+1} X_{n} 
- {\rm h.c.}\right] \right\rangle \;.
\eea
Using the Kubo formula and expanding to $O(t_c^2)$ gives the
conductivity, ie., the linear repsonse to an applied electric field in the 
$c$-direction $\sigma_c=\bar{j}^{c}/E_c$ :
\bea \label{sigma1}
\sigma_c(k_c=0,\Omega=0) & = & \lim_{\Omega \rightarrow 0} 
\frac{e ^2 t_c^2}{\Omega} 
\sum_n \int d^2 k_{\|} \int d \tau  \\
 &  & \times e^{i \Omega \tau}  U(\tau)
G^{(2d)}_n(\vec{k_{\|}}, \tau) G^{(2d)}_{n+1}(\vec{k_{\|}}, \tau) , \nonumber\\
U(\tau) & = & \left\langle X^{\dagger}_{n+1}(\tau) X_{n}(\tau) 
X^{\dagger}_{n}(0) X_{n+1}(0) \right\rangle_{H_{\rm ph}} . \nonumber
\eea where $ G^{(2d)}_n(\vec{k_{\|}})$ is the in-plane (dressed) Green's
function for the plane $n$ at in-plane momentum $\vec{k_{\|}}$.
In Fourier space, the effect of strong electron-phonon coupling
is to induce an effective $\omega$-dependent vertex $t_c^2 \longrightarrow
t_c^2 U(\omega)$. Classically, the phonon correlation function $U(\omega)$
leads to a Debye-Waller factor.

For magnetoresistance, we shall only study the case where
the magnetic field lies in-plane, 
avoiding the complication of magnetic field induced orbital 
effects on the in-plane propagators. We also do not consider spin
effects  here, see Discussion. As  $\vec{B}$ does not couple directly
to the phonons, its sole effect is on the orbital motion of the
electrons, which can be dealt with by the usual Peierls substitution
$\vec{p} \longrightarrow \vec{p} - e \vec{A}$.
With $\vec{B} = (B,0,0)$, the associated vector potential (in the
Landau gauge) is $\vec{A} = (0, -z B,0)$. Choose plane $n=1$ to have
$c$-axis coordinate $z=0$, and for plane $n=2$, $z=c$ where 
$c$ is the interplane distance. Hence,
\bea
G^{(2d)}_{n=1}(\vec{k_{\|}}) & \longrightarrow & G^{(2d)}_{n=1}(\vec{k_{\|}}),
 \nonumber \\
G^{(2d)}_{n=2}(\vec{k_{\|}}) & \longrightarrow &
G^{(2d)}_{n=2}(\vec{k_{\|}} + \vec{q}_B) , \\
\vec{q}_B & = & e \vec{c} \times \vec{B}  , \label{vecq}
\eea where $\vec{c}$ is a vector of length $c$ pointing in the positive
$c$-direction, and $\vec{k_{\|}}$ is the in-plane momentum.

To massage the (magneto) conductivity into a more familiar form, we
introduce the usual spectral representations:
\bea
G^{(2d)}_n(\vec{k_{\|}}, \omega_m) & = & \int dz 
\frac{A^{(2d)}(\vec{k_{\|}}, z)}{i \omega_m - z}, \nonumber \\
U(\nu_n) & = & \int dz 
\frac{B(z)}{i \nu_n - z}  \nonumber .
\eea where $\omega_m = (2 m + 1) \pi T $ is an odd Matsubara frequency, 
and $\nu_n = 2 n \pi T$ is an even Matsubara frequency. 
$T$ is the temperature.
Note that since there is translation invariance in the $c$-direction, 
we have dropped the plane index $n$ in the spectral functions 
$A^{(2d)}(\vec{k_{\|}}, z)$ and $B(z)$.
Fourier transforming with respect to the imaginary time 
in Eq.~ (\ref{sigma1}), and  doing the Matsubara sums leads to
the d.c. conductivity:
\bea \label{sigmakey}
\sigma_c(T,\vec{q}_B) & = & \frac{e^2 N_c t_c^2}{T}
\int d\omega\; d\nu \int  \frac{d^2 k_{\|}}{(2 \pi)^2} \\
 &  \times & A^{(2d)}(\vec{k_{\|}} + \vec{q_B}, \omega+\nu) 
A^{(2d)}(\vec{k_{\|}}, \nu) \nonumber \\
& \times & \left[ \frac{f(\nu) - f(\omega+\nu)}{\omega} \right]
D(\omega,T) ,\nonumber \\
D(\omega,T) & =  & \omega  \;
 B(\omega) n^B(\omega) \left[1+n^B(\omega)\right]  .\nonumber 
\eea where $N_c$ is the number of planes in the $c$-direction, 
 $f(\nu)=1/(\exp[\nu/T] +1)$ is the Fermi function,
and $n^B(\omega)=1/(\exp[\omega/T] -1)$ is the Bose function. $D(\omega,T)$
is essentially the phonon spectral weight multiplied by the Bose factors. 

\subsection{Phonon Spectral Function}
To get the phonon spectral function $B(\omega)$, we need:
\bea \label{phononspect}
B(\omega) = -\frac{1}{\pi} {\rm Im}\; U^{\rm ret}(\omega) = 
-\frac{1}{\pi} {\rm Im}\; \int_{-\infty}^{+\infty} e^{i \omega t} U^{\rm ret}(t) ,
\eea where the retarded real time spectral function is just:
\bea
U^{\rm ret}(t) = -i \theta(t) \left\langle \left[ 
X^{\dagger}_{n+1}(t) X_{n}(t) \; , \; 
X^{\dagger}_{n}(0) X_{n+1}(0) \right] \right\rangle_{H_{\rm ph}}
\eea with $X_n(t)$ given in Eq.~\ref{H_can}.

The calculation of $U^{\rm ret}(t)$ proceeds along the same line as in
the small polaron model (see Mahan\cite{Mahan}, Ch. 6.2), and
we get the exact result:
\bea \label{Uret}
U^{\rm ret}(t)  =  -i \theta(t) 2 
e^{\sum_q F_q [n^B(\omega_q) 
(1+ n^B(\omega_q)) -(2 n^B(\omega_q) +1)]} \nonumber \\
\qquad \quad  \times   \left\{ \cos\left[\omega_q (t+ i/2T)\right]
 - \cos\left[\omega_q (-t + i/2T)\right] \right\},
\eea where $F_q = \left|\frac{M_q}{\omega_q}\right|^2 2 (1-\cos q) > 0$,
and $n^B(\omega_q) = 1/(\exp(\omega_q/T) -1)$. 

In this paper, we look at two types of phonon dispersion: (1) Einstein
phonon where $\omega_q = \omega_0$ for all $q$, and (2) a generic dispersion
such that there are no sharp features in the phonon density of state.
Again, the calculation is directly analogous to that of the small
polaron problem, see e.g. chapter 4 of Ref.\cite{Mahan} for details of
the calculation.

For the Einstein phonon, setting  $\omega_q = \omega_0$ in Eq.~\ref{Uret},
putting this form into Eq.~\ref{phononspect},
and using the Bessel function identity $\exp(z \cos(\phi)) = 
\sum_{n=-\infty}^{+\infty} I_n(z) \exp(i n \phi)$, the spectral function
can be expressed as an infinite  sum of Bessel functions of the imaginary
kind:
\bea \label{B-Eins}
B(\omega) & = & 2 e^{-2 S_T} \\
& & \times \sum_{n=-\infty}^{+\infty} I_n(\Lambda(T)) 
\sinh \left( \frac{n \omega_0}{2 T} \right) \delta(\omega - n
\omega_0) ,  \nonumber
\eea  where 
\bea
\Lambda(T) &  = &  (\Delta/\omega_0)^2 /\sinh(\omega_0/2 T) ,\\ 
\Delta^2 & = &\sum_q |M_q|^2 2 (1-\cos q)  , \nonumber \\
 2 S_T & = & (\Delta/\omega_0)^2 \coth (\omega_0/2 T) . \nonumber
\eea Thus the function $D(\omega,T)$ defined in
Eq.~\ref{sigmakey} becomes:
\bea \label{DE}
D(\omega,T)  =  e^{-2 S_T} \Bigg\{ T I_0(\Lambda) \delta(\omega) \qquad \qquad 
\qquad \qquad  \\
\quad  +  \frac{\omega}{2 \sinh (\omega/2 T)} \sum_{n=1}^{\infty} I_n(\Lambda) 
\left[\delta(\omega - n \omega_0) + \delta(\omega + n
\omega_0)\right]\Bigg\}. \nonumber
\eea

For the more general dispersion, the 
time integral has to be approximated.  As in the small polaron problem,
one appeals to the fact that this is at strong electron-phonon coupling,
and so the dimensionless factor $F_q$ in the exponential in Eq.~\ref{Uret} is
large, and the integral can be done using the saddle
point approximation:
\bea \label{B_gen}
B(\omega) & \simeq & \frac{1}{\sqrt{\pi \gamma^2}} 
\sinh(\omega/2 T) e^{-\omega^2/4\gamma^2}  \\
&   \times & \exp\left\{-\sum_qF_q
\left[\sqrt{n^B(\omega_q)+1} -\sqrt{n^B(\omega_q)}\right]^2\right\} 
  ,\nonumber
\eea with $\gamma^2 = \sum_q |M_q|^2 (1-\cos q)/\sinh(\omega_q/2 T)$
and $F_q$ is as before (see after Eq.~\ref{Uret}). And the function
$D(\omega,T)$ becomes:
\bea \label{DG}
D(\omega,T) & \simeq & \frac{1}{4\sqrt{\pi \gamma^2}} 
\frac{\omega}{\sinh(\omega/2 T)} e^{-\omega^2/4\gamma^2} \\
 & \times & \exp\left\{-\sum_qF_q
\left[\sqrt{n^B(\omega_q)+1} -\sqrt{n^B(\omega_q)}\right]^2\right\}
 .\nonumber
\eea

\subsection{Calculation of $\sigma_c$} 
It is advantageous to do the in-plane momentum integrals first.
Note that the ratio of the magnetic wavector 
$\vec{q}_B = e \vec{c} \times \vec{B}$ 
(Eq.~\ref{vecq}) to the in-plane Fermi momentum $k_{\rm F}$ is:
\bea
\frac{q_B}{k_{\rm F}} \sim \frac{e c B}{\pi/2 a} = \frac{2 a c B}{\Phi_0}
\eea where $\Phi_0 = h/2e$ is the unit flux quantum, $a$ is the
in-plane lattice spacing, $c$ is the interlayer distance.
For $\rm Sr_2 Ru O_4$, 
$a=3.87 {\rm \AA} \;\; c=6.37 {\rm \AA}$ 
around room temperature, and so,
\bea
\frac{q_B}{k_{\rm F}} \sim 3 \cdot 10^{-4} \cdot B 
\quad\mbox{({\it B} in tesla)}. 
\eea Thus for the in-plane electronic dispersion, we can approximate 
$\epsilon_{\vec{k}+\vec{q}_B} \simeq \epsilon_{k} 
+ \vec{v} \cdot \vec{q}_B$ with
$\vec{v} = \partial \epsilon_{\vec{k}} /\partial \vec{k}$. 

We simplify further
with the choice of a flat band of width $2D$ 
and a circular (in-plane) Fermi surface. Then 
$\int d^2 k \longrightarrow N_0 \int_{-D}^{D} d\epsilon \int_{0}^{2\pi}
\frac{d\theta}{2\pi}$, where $N_0$ is the $2d$ density of state at
the Fermi surface. For the isotropic in-plane Fermi surface,
we pick $\vec{q}_B$ to lie in the direction of $\hat{k}_x$,
and thus, 
$\epsilon_{\vec{k}+\vec{q}_B} \simeq \epsilon_{k}+v_{\rm F}
q_B\cos(\theta)$.
For simplicity, we have taken $\vec{B}$ to lie entirely in the plane,
hence $q_B = e c B$.

Write the $2d$ spectral weight in the form:
\bea
 A^{(2d)}(\vec{k}, \omega) = \frac{1}{\pi} 
\frac{\Gamma(\omega)}{(\omega - \epsilon_{\vec{k}})^2 + \Gamma(\omega)^2} ,
\eea where we assume that the scattering rate $\Gamma(\omega)$ has
little in-plane momentum dependence. 
As all energy scales will be small compared to the bandwidth $2D$,
we can set $D\longrightarrow \infty$, and we find then,
\bea \label{Aint}
\int d^2 k A^{(2d)}(\vec{k}+\vec{q}_B, \omega + \nu) A^{(2d)}(\vec{k},
\omega) \qquad \qquad  \\
\qquad \simeq  \frac{N_0}{\pi} {\rm Re}  
\left[(v_{\rm F} q_B)^2 - (\nu + i \Gamma_{\rm tot}(T,\omega,\nu))^2\right]^{-1/2} ,\nonumber
\eea where $ \Gamma_{\rm tot}(T,\omega,\nu) = \Gamma(T,\omega) + 
\Gamma(T,\omega+\nu)$.

Substituting Eq.~\ref{Aint} into Eq.~\ref{sigmakey} leads finally to
the zero-frequency, zero-momentum $c$-axis conductivity:
\bea \label{SIGMAr}
\sigma_c(T,B) & = & \frac{e^2 N_0 N_c t_c^2}{\pi T} 
 {\rm Re} \int d\omega d\nu \frac{ f(\omega)-f(\omega+\nu)}{\nu} \nonumber \\
&   \times &
\frac{D(\nu,T)}{\left[(v_{\rm F} q_B)^2 -(\nu + i \Gamma_{\rm tot}(T,\omega,\nu))^2\right]^{1/2}}  .
\eea

This is the key equation we shall analyse in the next section. It is worth
summarizing the assumptions used in deriving this: 
(i) $\vec{B}$ lies in-plane and is small, 
(ii) flat band, circular Fermi surface, (iii) in-plane
spectral weight has a scattering rate that has little momentum dependence,
and (iv) second order perturbation theory in $t_c$.

\section{Results}

\subsection{Einstein phonons}

Because of the totally degenerate spectrum $\omega_q = \omega_0$,
the function $D(\omega,T)$  (Eq.~\ref{DE}) is made up of
delta funtions at the harmonics  
$\omega = n \omega_0, \; n = 0, \pm1, \pm2, \ldots$.
Putting this form into Eq.~\ref{SIGMAr} for a general in-plane
scattering $\Gamma(T,\omega)$, we get the exact result:
\begin{widetext}
\bea \label{Einsexact}
\sigma_c(T,B)  & = & \frac{ e^2 N_0 N_c t_c^2}{\pi} e^{-2 S_T} \int d\nu  
 \Bigg\{ \frac{-\partial f}{\partial \nu} \frac{a_0(\Lambda(T))}{\sqrt{(v_{\rm F}
e c B)^2 + (\Gamma_{\rm tot}(T,0,\nu))^2}} \\
 &+ &  \sum_{n=1}^{\infty} \bigg[ a_n(\Lambda(T)) 
\frac{f(\nu)-f(\nu + n\omega_0)}{n \omega_0} {\rm Re}
 \frac{1}{\sqrt{(v_{\rm F} e c B)^2 + 
(\Gamma_{\rm tot}(T,n\omega_0,\nu) - i n\omega_0 )^2}} 
 +   (\omega_0 \longrightarrow - \omega_0) 
\bigg]  \Bigg\} \nonumber ,
\eea \end{widetext}  where $a_0(\Lambda)=I_0(\Lambda), 
a_n(\Lambda)=I_n(\Lambda)\frac{n\omega_0/2T}{\sinh(n\omega_0/2T)}$, and $I_n(z)$ are
the Bessel functions of imaginary kind. One can show that the
amplitude, $a_n$, decreases with increasing $n$. 
In the experimentally accessible range of temperatures (at most
a few times of $T_{Debye} \sim 500 K$), one only needs
the first few harmonics ($n$) in Eq.~(\ref{Einsexact}).
This is because the higher harmonics contribute little to the
low $T$ and small $B$ behaviour, thanks to both the amplitude
$a_n(\Lambda)$ getting smaller and the denominator of Eq.~\ref{Einsexact}
getting larger with higher $n$. 

To determine the asymptotes, note that the various parameters of the 
Einstein phonon model have the limiting values (see after 
Eq.~\ref{B-Eins} for their definitions):
\bea
 \Lambda(T) & \longrightarrow & \left\{ \begin{array} {r@{\quad:\quad}l}
\left(\frac{\Delta}{\omega_0}\right)^2 \frac{2 T}{\omega_0}
 & T \gg \omega_0 , \\ 
\left(\frac{\Delta}{\omega_0}\right)^2 2 e^{-\omega_0/2T} &  T \ll
\omega_0 ,
\end{array} \right. \nonumber 
\eea
\bea 
2 S_T & \longrightarrow & \left\{ \begin{array} {r@{\quad:\quad}l}
\left(\frac{\Delta}{\omega_0}\right)^2 \frac{2 T}{\omega_0}
 & T \gg \omega_0 , \\  
\left(\frac{\Delta}{\omega_0}\right)^2 & T \ll \omega_0 ,
\end{array}\right.   \label{Einasympt}
\eea where $\Delta^2 = \sum_q |M_q|^2 2 (1-\cos q)$ characterises the
strength of the electron-phonon interaction. The argument $\Lambda(T)$ of
the Bessel function is always small in the low $T$ limit, but it
can be large or small in the high $T$ limit, depending on the magnitude
of $\Delta/\omega_0$ via the combination $(\Delta/\omega_0)^2 (T/\omega_0)$.

Substituting the asymptotic forms Eq.~\ref{Einasympt} into Eq.~\ref{Einsexact},
and using the expansions for Bessel function,
we get for low temperatures $T\ll\omega_0$, (or for high $T$ with 
$(\Delta/\omega_0)^2 (T/\omega_0) \ll 1$):
\bea \label{Einslo}
\sigma_c & \simeq & \frac{ e^2 N_0 N_c t_c^2}{\pi}
  \frac{ e^{-(\Delta/\omega_0)^2}  }{\sqrt{(v_{\rm F} e c B)^2
+ 4 \Gamma(T,0)^2}} .
\eea  We have dropped $n>0$ terms, as they are of order $e^{-n
\omega_0/T}$ relative to the $n=0$ term. 
Also, both the marginal Fermi liquid and the Fermi liquid
scattering rate varies relatively slowly compared to $\partial
f/\partial \nu$, and hence we have approximated $-\partial
f/\partial \nu \sim \delta(\nu)$ to get the above equation. 
Note that in this low $T$ limit, the 
effective bandwidth $t_c$ gets renormalised by the phonons by the
factor $\exp [-(\Delta/\omega_0)^2 /2]$.

For high temperatures $T\gg\omega_0$ 
 and $\Delta/\omega_0 > 1$:
\bea \label{Einshi}
\sigma_c  & \simeq & \frac{ e^2 N_0 N_c t_c^2}{\pi}
\frac{1 + O(\omega_0/T)}{\sqrt{2 \pi \left(\frac{\Delta}{\omega_0}\right)^2
\frac{2 T}{\omega_0}}}  \\
& \times & \Bigg\{ \frac{1}{\sqrt{(v_{\rm F} e c B)^2 + 4
 \Gamma(T,0)^2}}  \nonumber \\
 & & + 2 {\rm Re} \frac{1}{\sqrt{(v_{\rm F} e c B)^2 + 
(2 \Gamma(T,\omega_0) -i \omega_0)^2}} \nonumber  \\
& & +  {\rm (higher\;\; harmonics)} \Bigg\} .
\nonumber
\eea Again, we have made the approximation $(f(\nu)-f(\nu+n\omega_0))/n
\omega_0 \sim -\partial f/\partial \nu \sim \delta(\nu)$, the first
approximation being valid as $T\gg \omega_0$, and the second
approximation is due to the
same reason as mentioned for the low temperature case.
Notice that the leading terms in both the high- and low-temperature
limits do not have any $e^{-\Delta/T}$ factors, in contrast to the
general phonon dispersion case below. 

The precise low and high $T$-
dependence at $B=0$ reflects the form of the in-plane scattering. For a
marginal Fermi liquid form:
\bea \label{MFL}
\Gamma(T,\omega) = \alpha_{\rm MFL} \omega \coth\left(\omega/T\right),
\eea the asymptotics from Eqns.\ref{Einslo},\ref{Einshi} are:
\bea
\sigma_c(T\ll\omega_0) & \sim  & T^{-1} \\
\sigma_c(T\gg\omega_0) & \sim  & T^{-3/2} \nonumber
\eea while for a Fermi liquid form:
\bea \label{FL}
\Gamma(T,\omega) = \alpha_{\rm FL} \frac{\omega^2 + (\pi \; T)^2}{\epsilon_{\rm F}} ,
\eea the asymptotics are:
\bea
\sigma_c(T\ll\omega_0) & \sim  & T^{-2} \\
\sigma_c(T\gg\omega_0) & \sim  & T^{-5/2} \nonumber
\eea

\begin{figure}[h]
\includegraphics[width=\columnwidth]{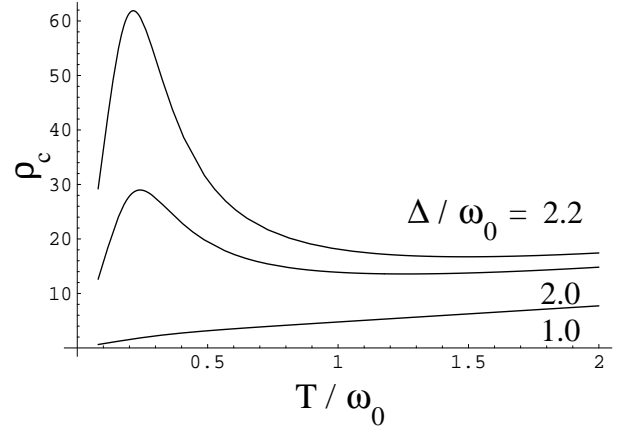}
\caption{$B=0$  $c$-axis resistivity  with in-plane marginal Fermi 
liquid scattering (Eq.~\ref{MFL}) for electrons coupled to $c$-axis 
Einstein phonons, $\alpha_{\rm MFL}=1.0$}
\label{inter2}
\end{figure}

\begin{figure}[h]
\includegraphics[width=\columnwidth]{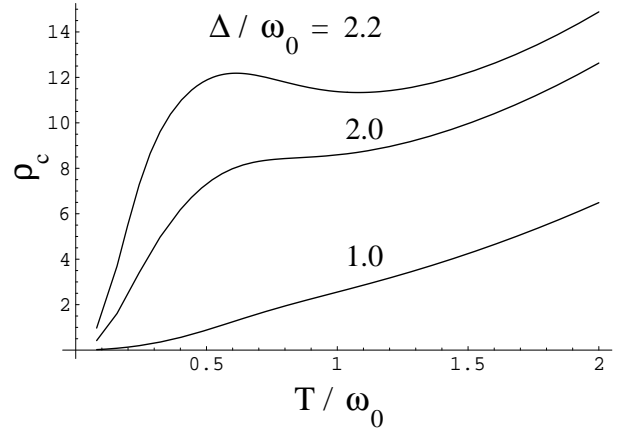}
\caption{$B=0$ $c$-axis resistivity with in-plane Fermi 
liquid scattering (Eq.~\ref{FL}) for electrons coupled to $c$-axis Einstein phonons,
 $\alpha_{\rm FL}=5.0$,
 $\epsilon_{\rm F}=100\omega_0$. }
\label{inter3}
\end{figure}

The above asymptotics miss out the experimentally important and
interesting regime of $T\approx\omega_0$. A broad maximum appears when
$T$ is somewhat below $\omega_0$, {\it only} when $\Delta/\omega_0
\gtrsim 2$.
Also, the  maximum grows rapidly with increasing $\Delta/\omega_0$.
This occurs because of two  competing  factors: see Eq.~\ref{Einsexact}.
For $\Delta/\omega_0 \gg1$, $\Lambda(T)$ is again large, and we can use the
asymptotic form for the Bessel functions, while keeping $T\approx\omega_0$:
\bea \label{Einshump}
\sigma_c & \propto & \sqrt{\sinh(\omega_0/2T)}
\exp\left(-\left[\frac{\Delta}{\omega_0}\right]^2
2 \frac{\sinh^2(\omega_0/4T)}{\sinh(\omega_0/2T)}\right) \nonumber \\
& & \times
\left\{ \frac{1}{\left[(v_{\rm F} e c B)^2 + (2\Gamma(T,0))^2\right]^{1/2}} 
+ \cdots \right\} , 
\eea where we have dropped higher harmonics (which change the magnitude,
but not the general $T$-dependence). Thus there are two opposing
tendencies: the phonon contributed part (the exponential and the square
root of sinh) which grows with $T$, versus the in-plane scattering
part $\Gamma(T,0)\propto T$ for the marginal Fermi liquid. 
The result is the broad maximum structure in
$\sigma_c$  (Fig.~\ref{inter2},\ref{inter3}).
 Note that the height of the  maximum has a strong dependence
on the strength of the electron-phonon coupling $\Delta$.
In a Fermi liquid the  in-plane scattering rate $\sim T^2$ damps out the phonon
contributed parts more strongly and to see this structure, one needs 
to go to larger  $\Delta/\omega_0$. 

We have estimated the critical
phonon parameter $\Delta_c/\omega_0$ and the position of the  maximum
$T_{\rm max}$ and the concomitant minimum $T_{\rm min}$ 
from Eq.~\ref{Einsexact} by taking into account only
the zeroth harmonics $n=0$. For an in-plane scattering rate of the
form $\Gamma(T) = \alpha T^{\eta}$,  
\bea  \label{Delta_c}
\frac{\Delta_c}{\omega_0} \approx  \sqrt{\frac{\eta}{0.4}} ,
\eea
and the broad maximum only appears when $\Delta > \Delta_c$.
The critical temperature when the broad maximum starts to appear,
{\it i.e.} when $T_{\rm max}$ starts to be different from $T_{\rm min}$,
occurs at:
\bea
T_{\rm crit} \approx  \omega_0 / 3.0  .
\eea For $\Delta$ only slightly larger than $\Delta_c$, 
both $T_{\rm max}$ and $T_{\rm min}$ are close to $T_{\rm crit}$:
\bea
\frac{\omega_0}{2 T_{\stackrel{\scriptstyle \rm max}{\rm min}}} 
\approx \frac{\omega_0}{2 T_{\rm crit}} \pm  \sqrt{\frac{\eta}{0.2}
\left[\left(\frac{\omega_0}{\Delta_c}\right)^2 
-\left(\frac{\omega_0}{\Delta}\right)^2 \right]}  . 
\eea
For large $\Delta/\omega_0$, $T_{\rm max}$ and $T_{\rm min}$ 
obey the approximate equations:
\bea  \label{Thump}
\frac{\Delta^2}{\omega_0^2} & \approx & \frac{\eta T_{\rm max}}{\omega_0} \exp
\left(\frac{\omega_0}{2 T_{\rm max}}\right) ,\\
\frac{\Delta^2}{\omega_0^2} & \approx & \frac{4 \eta T_{\rm
min}}{\omega_0} . \nonumber
\eea  
Note that both $T_{\rm max}$ and $\Delta_c$
depend mainly on phonon parameters; one can show
that the in-plane scattering rate enters only in the 
form of the exponent $\eta$. In particular, 
both  $T_{\rm max}$ and $T_{\rm min}$ can only depend on the scale
$\omega_0$. Hence the width of the broad maximum, roughly given by
 $2 (T_{\rm max}-T_{\rm min})$, is governed by the scale $\omega_0$.

Fig.~\ref{inter2} illustrates the resistivity (the inverse of the 
conductivity) of
Eq.~\ref{Einsexact} for a range of values of $\Delta$, using 
a marginal Fermi liquid form for the in-plane scattering, and including up
to $n=5$ harmonics of $\omega_0$. Fig.~\ref{inter3} shows the Fermi
liquid case, where the maximum is much less prominent than in the
marginal Fermi liquid case, as already mentioned.\cite{FLnotes}
All energies are measured in units of $\omega_0$. $\rho_c$ is
measured in units of $\pi/e^2 N_0 N_c t^2_c$.

\begin{figure}[h]
\includegraphics[width=\columnwidth]{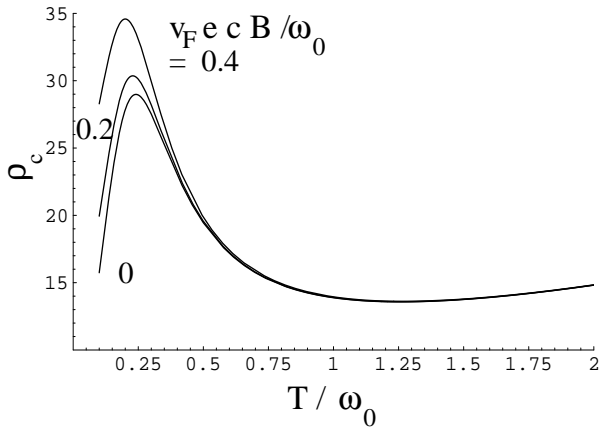}
\caption{$c$-axis resistivity at finite $B$ with 
in-plane marginal Fermi 
liquid scattering (Eq.~\ref{MFL}), for electrons coupled to $c$-axis 
Einstein phonons as a function of $T$. 
$\alpha_{\rm MFL}=1.0$, $\Delta/\omega_0 = 2.0$.}
\label{BT-MFL}
\end{figure}

\begin{figure}[h]
\includegraphics[width=\columnwidth]{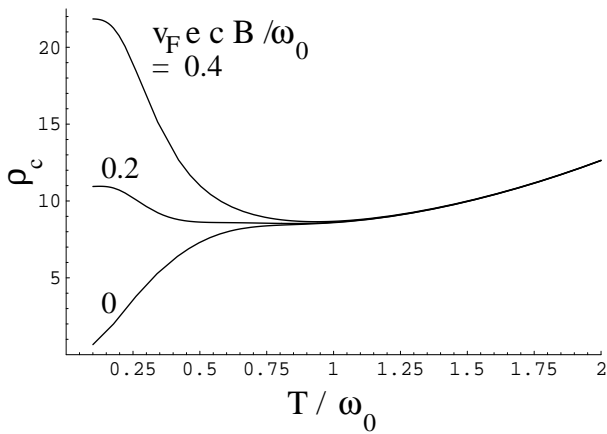}
\caption{$c$-axis resistivity at finite $B$ with in-plane Fermi 
liquid scattering (Eq.~\ref{FL}), for electrons coupled to $c$-axis Einstein phonons, 
as a function of $T$. 
$\alpha_{\rm FL}=5.0$, $\epsilon_{\rm F}=100\omega_0$, $\Delta/\omega_0 = 2.0$. }
\label{BT-FL}
\end{figure}

Now we consider the magnetic field dependence. The magnetoresistivity
is defined as $\Delta \rho_c = \rho_c(B)/\rho_c(0) -1$.  For later 
convenience, the  effective cyclotron frequency is defined as:
$\omega_c =  v_{\rm F} e c B$  .

\begin{figure}[h]
\includegraphics[width=\columnwidth]{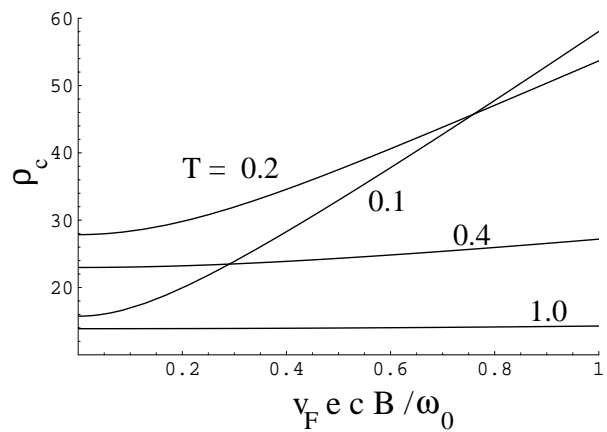}
\caption{$c$-axis resistivity with in-plane marginal Fermi 
liquid scattering (Eq.~\ref{MFL}), for electrons coupled to  $c$-axis Einstein
phonons,  as a function of $B$. 
$\alpha_{\rm MFL}=1.0$, $\Delta/\omega_0 = 2.0$. $T$ is measured in
units of $\omega_0$.}
\label{B-MFL}
\end{figure}

\begin{figure}[h]
\includegraphics[width=\columnwidth]{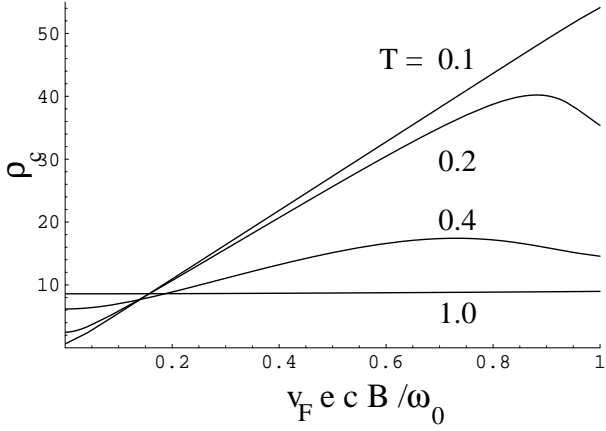}
\caption{$c$-axis resistivity with in-plane Fermi 
liquid scattering (Eq.~\ref{FL}), for electrons coupled to  $c$-axis Einstein
phonons, as a function of $B$. 
$\alpha_{\rm FL}=5.0$, $\epsilon_{\rm F}=100\omega_0$, 
$\Delta/\omega_0 = 2.0$. $T$ is measured in units of $\omega_0$.}
\label{B-FL}
\end{figure}

 First, at very low $T$ and  very large magnetic fields,  
in Eq.~\ref{Einslo} the  effective cyclotron frequency 
can become larger than the scattering rate 
and so the magnetoresistance goes as:
\bea \label{MR-BigB}
\Delta \rho_c \propto \omega_c /\Gamma(T) ,
\eea {\it i.e.}, {\it linearly}\cite{SchofieldCooper} in $B$.
See Fig.~\ref{B-MFL},\ref{B-FL}.

At higher $T$, the scattering rate dominates in both 
Eqs.~\ref{Einshi} and \ref{Einshump}. For the $n=0$ term, 
$[\omega_c^2 + \Gamma^{\rm tot\;\;2}]^{-1/2}$
gives a $1 - B^2$ contribution to the conductivity, {\it i.e.}
a $+B^2$ magnetoresistance. For the $n \geq 1$ terms,
since $\omega_0 \sim 10^{-2} \epsilon_{\rm F}$, and 
$\omega_c \sim 10^{-3} \epsilon_{\rm F}$ for $B$ up to a few tesla,
$\omega_c \ll \omega_0$, we can extract the leading
$B$ behaviour of the $n \geq 1$ terms in the conductivity:
\bea \label{Bbehav}
{\rm Re} \frac{1}{\sqrt{\omega_c^2 + 
(\Gamma_{\rm tot} -i n \omega_0)^2}} \simeq
\frac{\Gamma_{\rm tot}}{\Gamma_{\rm tot}^2 + (n \omega_0)^2} \qquad
\quad\\
\times \left( 1 - \omega_c^2 \frac{\Gamma_{\rm tot}^2 - 3 (n \omega_0)^2}{2
\left[\Gamma_{\rm tot}^{2} + (n \omega_0)^2\right]^2} + O(\omega_c /
\Gamma_{\rm tot})^4 \right) . \nonumber
\eea While it is possible for the coefficient of the $B^2$ 
term in Eq.~\ref{Bbehav}
to change sign, it can be shown that the $n=0$ term always dominate,
leading to a $1- B^2$ conductivity, or a $+B^2$ magnetoresistance: 
\bea  \label{MR-eins}
\Delta \rho_c \propto (\omega_c /\Gamma_{\rm eff})^2  ,
\eea 
where the effective scattering rate $\Gamma_{\rm eff}$ is made up of
both the electronic scattering and the phonon frequency, and
taking into account only up to $n=1$ harmonics,
\bea
\Gamma_{\rm eff}^2 \approx 
\frac{\frac{1}{2\Gamma_0} +\frac{4 \Gamma_1}{4 \Gamma_1^2 +\omega_0^2}}
{\frac{1}{16 \Gamma_0^3} +\frac{2 \Gamma_1 (4 \Gamma_1^2 - 3
\omega_0^2)}{(4 \Gamma_1^2 +\omega_0^2)^3}}  ,
\eea where $\Gamma_0=\Gamma(T,0)$, and $\Gamma_1=\Gamma(T,\omega_0)$.

In conclusion, there is positive magnetoresistance
at all temperatures and for both Fermi liquid and marginal Fermi
liquid in-plane states. In general, if the in-plane scattering
$\Gamma(\omega)$ could be made {\it larger} at $\omega=0$ relative
to the higher harmonics $ \omega= n \omega_0$, the contribution
of the $n=0$ part of the phonon  function $D(\omega)$ is suppressed,
and this can lead to a {\it negative} magnetoresistance. 

Figs.~\ref{BT-MFL},\ref{BT-FL},\ref{B-MFL} and \ref{B-FL} illustrate
the resistivity as a function of $T$ and $B$.

\subsection{General phonon dispersion}

Having considered an Einstein mode, we now treat a more general 
dispersing phonon mode.
The function $D(\omega,T)$ will be seen to be strongly peaked near
$\omega \sim 0$, compared to the in-plane Fermi energy scale $\epsilon_{\rm F}$.
 Furthermore, we are in the regime $T/\epsilon_{\rm F} \ll 1$. Thus, 
$[f(\nu)-f(\nu+\omega)]/\omega \simeq -\partial f/\partial \nu 
\simeq \delta(\nu)$. The
last approximation is valid for the same reason as in the Einstein
phonon case. Hence, with these approximations and
inserting the gaussian form of the phonon function of Eq.~\ref{DG} into
Eq.~\ref{SIGMAr} leads to
\bea \label{siggen}
\sigma_c(T,B) & = & \frac{e^2 N_0 N_c t_c^2}{\pi T} 
e^{-C(T)} \int \frac{d\nu}{\sqrt{\pi \gamma^2(T)}} 
\frac{\nu}{\sinh \frac{\nu}{2 T}} \nonumber \\
& & \times{\rm Re}
 \frac{\exp-\frac{\nu^2}{4\gamma^2(T)}} 
 {\left[(v_{\rm F} e c B)^2 - (\nu + i \Gamma_{\rm
tot}(T,\nu))^2\right]^{1/2}} \nonumber \\
C(T) & = & \sum_q F_q 
\frac{2 \sinh^2 \left(\frac{\omega_q}{4 T}\right)}
{\sinh\frac{\omega_q}{2 T}} 
\eea where $F_q =  \left|\frac{M_q}{\omega_q}\right|^2 2 (1-\cos q)$ and
$\gamma^2(T) = \sum_q F_q |\omega_q|^2 /2 \sinh(\omega_q/2T)$.

The various parameters have the following limiting values:
\bea \label{genhiT}
\gamma^2 (T) & \longrightarrow & \left\{ \begin{array} {r@{\quad:\quad}l}
2 T \bar{\Delta} & T \gg \omega_q ,\\ 
\gamma^2_0(T)  &  T \ll \omega_q ,
\end{array} \right. \nonumber \\
\exp-C(T) & \longrightarrow & \left\{ \begin{array} {r@{\quad:\quad}l}
e^{-\bar{\Delta}/T} & T \gg \omega_q ,\\  
e^{-\sum_q F_q} & T \ll \omega_q .
\end{array}\right. .
\eea   $\bar{\Delta} = \sum_q F_q \omega_q /2$ involves the
 electron-phonon coupling, together with the
phonon dispersion. It characterises the strength of the electron-phonon
interaction, and is  directly analogous to the parameter $\Delta$
for the Einstein phonon case.
$\gamma_0^2(T)= \sum_q F_q |\omega_q|^2 \exp-\omega_q/2T $. (Note that
the expressions $T\gg \omega_q$ and $T\ll \omega_q$ refer to some
typical  energy of the phonon dispersion, {\it e.g.} the Debye frequency.)

In the low temperature limit $T \ll \omega_q$, the phonon spectral
function provides a gaussian in frequency with a
width $\gamma_0(T)$ which is exponentially
narrow in $T$. Thus in the integral over frequency ($\int d\nu$), only
$\nu \approx 0$ contributes. Furthermore, since $T \gg \gamma_0(T)$,
the scattering rate $\Gamma(T,\nu)$ can be taken at zero frequency,
for both a Fermi liquid and a marginal Fermi liquid. Then the integral
in Eq.~\ref{siggen} can be done immediately, 
leading to the low $T$ asymptotic form:
\bea \label{losiggen}
\sigma_c(T\ll\omega_q,B) & = & \frac{2 e^2 N_0 N_c t_c^2 \;\;\;
e^{-\sum_q F_q}}{\pi \sqrt{(v_{\rm F} e c B)^2 + 4 \Gamma(T,0)^2}} .
\eea This gives rise then to a low temperature
metallic resistivity that tracks
the in-plane scattering $\Gamma(T,0)$, and
a positive magnetoresistance with a leading $B^2$-dependence,
much as in the Einstein phonon case. 
Note that the hopping matrix element $t_c$ is effectively reduced
by the phonon-induced  prefactor $ \exp-\sum_q F_q/2$, similar to the
Einstein phonon case.
 
In the high  temperature limit $T \gg \omega_q$, when $\bar{\Delta} \ll T$,
the width of the gaussian $\gamma^2 \sim 2 T \bar{\Delta} \ll T^2$, hence
again only small frequency $\nu \sim 0$ is needed in the integral in 
Eq.~\ref{siggen}, leading to:
\bea \label{hisiggen}
\sigma_c(T\gg\omega_q,B) & = & \frac{2 e^2 N_0 N_c t_c^2 \;\;\;
e^{-\bar{\Delta}/T}}{\pi \sqrt{(v_{\rm F} e c B)^2 + 4 \Gamma(T,0)^2}} .
\eea

\begin{figure}[h]
\includegraphics[width=\columnwidth]{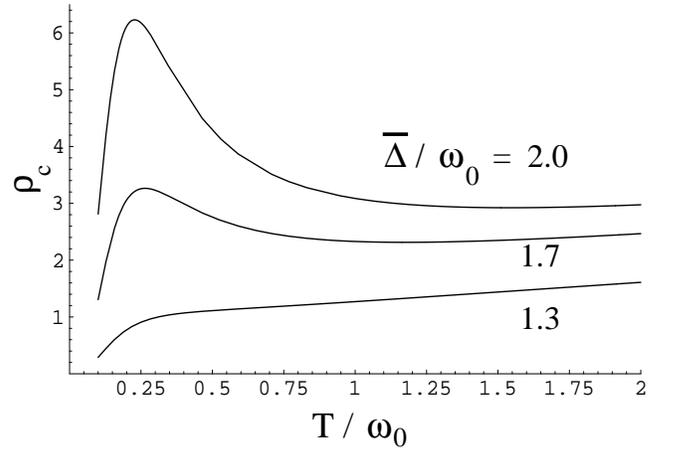}
\caption{B=0 $c$-axis resistivity with 
in-plane marginal Fermi  liquid scattering , for the phonon dispersion 
$\omega_q = \omega_0 [1- 0.4 \cos(q)] $, $\alpha_{\rm MFL}=1.0$.}
\label{fig4}
\end{figure}

\begin{figure}[h]
\includegraphics[width=\columnwidth]{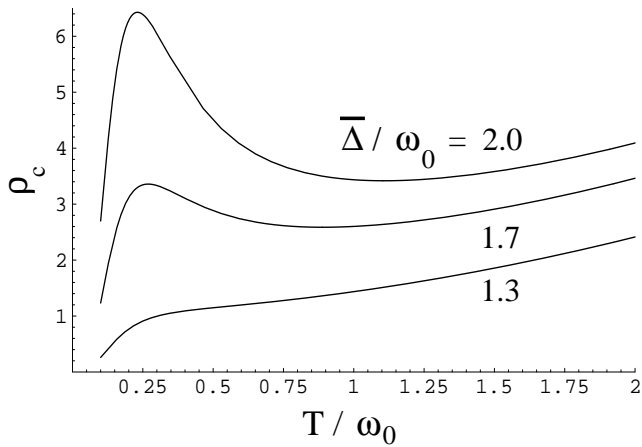}
\caption{B=0 $c$-axis resistivity with
in-plane  Fermi  liquid scattering , for the phonon dispersion 
$\omega_q = \omega_0 [1- 0.4 \cos(q)] $, $\alpha_{\rm FL}=5.0$, 
$\epsilon_{\rm F}=100\omega_0$. }
\label{fig5}
\end{figure}

Just as in the Einstein phonon case, there can be a broad maximum structure
when $T \sim \omega_q$: it can be shown that in this regime,
the phonon contribution $e^{-C(T)} /\gamma(T)$ provides 
 a semi-conducting-like upturn in conductivity, which competes with 
the in-plane scattering $\Gamma(T,0)$ that eventually dominates
as $T\gg \omega_q$. This is to some extent captured in 
Eq.~\ref{hisiggen} above.
More precise $T$-dependence will require an
explicit form for the optical phonon dispersion $\omega_q$.
For illustration, let
$\omega_q = \omega_0 [1- 0.4 \cos(q)] $, and set the electron-phonon coupling
$M_q$ to be $q$-independent. Eq.~\ref{siggen} is then evaluated 
numerically and plotted:
Fig.~\ref{fig4} is for  a marginal Fermi liquid form for the in-plane scattering,
while Fig.~\ref{fig5} shows the Fermi liquid case. $\rho_c$ is
measured in units of $\pi/ e^2 N_0 N_c t^2_c$. Qualtitatively, these plots are
similar to the Einstein phonon ones.

As for the Einstein phonon case, there is only positive magnetoresistance
for the more general phonon dispersion, at all temperatures.

\section{Discussion}

\subsection{Summary of results}

 In this paper we have studied $2d$ planes
of electrons characterised by a Fermi liquid or a marginal Fermi liquid
scattering rate (that has little momentum dependence), 
coupled by a small inter-plane single-particle hopping in the third
dimension,
in the presence of a strong interaction between electronic density
and a bosonic mode in the inter-plane
direction. The electron-boson coupling is treated exactly via a
canonical transformation, and we treat the
inter-plane hopping to lowest order ($t_c^2$) to calculate 
conductivity and magnetoconductivity. For concreteness, 
we have considered the bosonic mode to be a phonon, and have treated
two types of phonons: Einstein 
phonons and a more general dispersing phonon mode. The overall behavior
appears to be quite similar for the Einstein mode when
 compared to the more general
phonon dispersion, and we shall discuss them together, pointing
out the differences where they arise.

We find that at temperatures low compared to the typical phonon energy,
the $c$-axis resistivity $\rho_c$ is metallic, displaying the
same temperature dependence as that of the underlying 
in-plane scattering  rate $\Gamma(T)$:  $\rho_c \sim \Gamma(T)$. 
The magnitude is however strongly reduced, corresponding
to a much smaller effective $c$-axis bandwidth, directly analogous to
the bandwidth renormalisation in the Holstein small 
polaron problem\cite{Holstein,Mahan}. 

At temperatures  much much larger than the typical phonon energy,
(a regime that is probably academic,) $\rho_c(T)$
again reflects the $T$-dependence of the in-plane scattering 
rate, but in the Einstein phonon case, there is an extra multiplicative factor
of $T^{1/2}$ contributed by the phonons, while for the general phonon case,
the $T$- dependence is exactly as in  $\Gamma(T)$. Note that it is
because of this  in-plane factor of $\Gamma(T,\omega)$, that the high
temperature state is metallic like ({\it i.e.} $d\rho_c /dT > 0$), in
contrast to the standard Holstein small polaron model. 

In the experimentally interesting regime where $T$ is of the same
magnitude as the phonon energy (a few hundred Kelvins), we find the
possibility in
$\rho_c$ of a broad maximum structure  that connects the low 
and high temperature
metallic states. For the Einstein phonons, we have shown that this
broad maximum, {\it i.e.} a ``non-metallic'' $T$-dependence, occurs {\it only}
when there is strong electron-phonon coupling (Eq.~\ref{Delta_c}).
We have also estimated the position and
width of the maximum in the temperature axis.
For the general phonon dispersion case, we expect, and
see numerically, qualitatively similar behavior. 
Note that the physics is richer here than in
the Holstein small polaron problem, due to the extra degree of
freedom of the in-plane scattering rate.

As expected, we have found only positive magnetoresistance in our model, since
we only consider the effect of the magnetic field pointing in the
plane on the orbital motion of the electrons. This goes
as $\Delta \rho_c \propto (\omega_c /\Gamma_{\rm eff})^2$
(Eq.~\ref{MR-eins}),  except for the regime
when the magnetic field is very large and the temperature is
very low, in which case\cite{SchofieldCooper} 
$\Delta \rho_c \propto \omega_c /\Gamma(T)$ (Eq.~\ref{MR-BigB}).
 This is so for both the Einstein phonons and the more
general phonon dispersion, and for both the marginal Fermi liquid
or Fermi liquid scattering rates.  

In this work we have tacitly assumed that computing $\sigma_c$ as an
expansion in $t_c$ is valid and that we can fully characterize the
in-plane physics by its single particle properties. For non-Fermi
liquid states in the plane this may not be the case (a Luttinger
liquid in 1D is a notable example~\cite{bourbonnais}). 
Under such circumstances our
results would presumably breakdown below a temperature scale which
would depend on the details of the in-plane physics. 

Our model consists of phonons that have no structure in the
in-plane direction, and a more realistic model should have a fully
three dimensional dispersion. In this respect, a question immediately
arises: if the metallic to non-metallic transition in the $c$
direction requires a strong electron-phonon coupling, might not this
strong coupling also show up in the in-plane physics? This is
presumably a question of scales: the characteristic phonon energy
is expected to be larger than the $c$-axis hopping matrix element
$t_c$, but the phonon energy is still much smaller than the in-plane
Fermi energy, because of the highly anisotropic nature of the
materials considered here. Hence, one expects the strong
electron-phonon coupling to have a much more dramatic effect in
the $c$-direction, than in the in-plane direction.

\subsection{Application to the ruthenate systems}

There exist optical phonons with the
appropriate symmetry for $c$-axis transport\cite{Braden} in
 $\rm Sr_2 Ru O_4$, and experimentally
the broad maximum in the $c$-axis resistivity has been linked
to a structural phase transition\cite{Jin}
 in $\rm Ca_{1.7} Sr_{0.3} Ru O_4$ at around the broad maximum temperature.
Thus one should consider the possibility of electron-phonon interaction
affecting the $c$-axis transport.

In particular, both the $\rm Sr_2 Ru O_4$ and $\rm Ca_{1.7} Sr_{0.3} Ru O_4$ 
systems exhibit\cite{resist,Jin}
qualitatively this broad maximum structure found in our simple model,
in the $c$-axis resistivity 
near to their characteristic ($c$-axis)
phonon energy. In  $\rm Sr_2 Ru O_4$ there is even the
hint of an upturn in resistivity at  high temperatures\cite{upturn},
 and the result
plotted in Fig.~\ref{inter2} bears some resemblance to the data
of Tyler {\it et.al.}\cite{resist} (their Fig.~3). 
Also, the broad maximum temperature of $\sim 130$K in $\rm Sr_2 Ru O_4$
is indeed smaller than the characteristic phonon
energy (around 500 to 800 K,\cite{Braden}), 
consistent with our estimated bound $T_{\rm max} < 3 \omega_0$. Using
the value of 500K for $\omega_0$, we estimate that the 
electron-phonon coupling parameter $\Delta$ is about two times of $\omega_0$.

However, the negative magnetoresistance seen around the same temperature
regime as the broad maximum cannot at present be explained within this 
phonon-assisted hopping picture. 
In our simple model here, to obtain negative
magnetoresistance robustly, one needs either a phonon function
$D(\omega)$ that peaks at a {\it finite} frequency, and/or an in-plane
spectral weight where the scattering rate is smallest at a {\it finite}
frequency. Either of these possibilities suppress the $\omega=0$ contribution
to $\sigma_c$ (see Eq.~\ref{SIGMAr}.) The Kondo effect can give rise to the
second possibility, while it is unclear what kind of phonons can lead
 to the first possibility. 

Other factors that we have neglected that could influence 
the magnetoresistance include the details of the Fermi surface.  
We have  used  a flat 
band with a circular Fermi surface, while experimentally,
 the $\beta$ band (which  has  the largest $c$-axis 
dispersion\cite{dHvA}) is nearly a square Fermi surface and moreover, 
the $c$-axis electronic 
dispersion has dependence on the direction in the in-plane
momentum space. Also, the $\rm Ru O_2$ planes are staggered from one plane
to another, so that an electron hopping from one plane to another
does not stay at the same in-plane coordinate $\vec{x}$ as we have
assumed in Eq.~\ref{model}. 
These facts may lead to a more singular 
momentum-integrated spectral weight, compared to the form in Eq.~\ref{Aint}
used here. But if the spectral weight is now concentrated into a
narrower range of momenta, the $\vec{B}$-induced 
relative shift in the momenta  between the planes would lead to
a more positive magnetoresistance. 
  
Alternatively, there may be additional hopping channels
that contribute to the negative magnetoresistance, for example, hopping
via some intermediate localized state\cite{Hussey}.

Next, we discuss spin effects, which have been ignored so far.
At least in the low temperature Fermi liquid regime ($T<25K$),
there are  various types of spin fluctuations
in $\rm Sr_2 Ru O_4$ (both antiferromagnetic\cite{AFspin} and 
$q$-independent\cite{ferro}). In-plane spin fluctuations enhance
the in-plane electronic scattering rate. If the applied magnetic field 
freezes these fluctations, then one can get a negative magnetoresistance,
with the magnetic field pointing in any direction.
However, this negative magnetoresistance will be seen in both
$\rho_c$ and $\rho_{ab}$. In $\rm Sr_2 Ru O_4$, 
 $c$-axis magnetoresistance goes negative above $T_{\rm max}$
for magnetic fields pointing in-plane or in the $c$-direction,
but the in-plane magnetoresistance remains positive\cite{Hussey}.
Thus one can rule out suppression of in-plane scattering  due
to magnetic field freezing out spin fluctuations. And if there were
spin-fluctuation mediated $c$-axis hopping, magnetic field may again
suppress this, leading  to a positive magnetoresistance. 

After this work was completed, we learnt that the in-plane spectral weight
at temperatures near and above the resisitivity maximum temperature may be 
anomalous\cite{ARPES}, in contrast to the spectral weight observed in the
low temperature Fermi liquid state. In the Introduction, we have pointed to the possibility
that  anomalous in-plane physics can lead to incoherent $c$-axis
transport, and these recent results suggest that for   $\rm Sr_2 Ru O_4$, 
this needs to be taken into account. However, because of the linkage of the 
structural phase transition and the resistivity maximum,
the strong electron-phonon coupling model is still relevant to the anomalous $c$-axis
transport in $\rm Ca_{1.7} Sr_{0.3} Ru O_4$.

\subsection{Boson-electron coupling}

Any strong boson-electron coupling can in general be tackled by
the present technique. The analysis presented is for
the specific case of a neutral, gapped, bosonic mode, as in a phonon.
Instead of phonons, there could
be magnons or paramagnons describing the effects of spin fluctuations.
In addition to the cuprates, other members of the family of
strontium/calcium ruthenate exhibit some
types of spin fluctuations  and
work is in progress to check their importance in the $c$-axis transport. 
For magnetoresistance the calculation however needs to be modified. 
 In this paper, the effect of magnetic field is purely on the orbital motion
of in-plane electrons because 
there is no direct coupling between the magnetic field and the phonons,
which is not the case for (para)magnons.

\subsection{Conclusion} 

We have demonstrated that a strong coupling between electrons and
a bosonic mode in the (weakest) $c$-direction in a highly anisotropic
metal can lead to a broad maximum 
(metallic to non-metallic crossover) in the $c$-axis resistivity. For
the case where the boson is a phonon, we have
discussed the potential application of the model to $\rm Sr_2 Ru O_4$ and its
relative $\rm Ca_{1.7} Sr_{0.3} Ru O_4$. Despite certain simplifying
features of the model, the qualitative properties of this crossover in these
layered ruthenates are captured  succinctly. Work in progress
will address the issue of the anomalous $c$-axis magnetoresistance
in $\rm Sr_2 Ru O_4$ within this framework, and also the effect of the
anomalous in-plane spectral weight on $c$-axis transport in $\rm Sr_2 Ru O_4$. Also we
are currently studying other signatures in transport properties of this
strong electron-boson coupling, for example, in thermal conductivity.

After completion of this work, we became aware
of some related work by Lundin and McKenzie concerning a related small
polaron model for intra- and interlayer transport [cond-mat/0211yyy].
 We thank them for sending us a copy of their preprint prior
 to submission.

\vskip 0.5 truein

{\it Acknowledgement}  The authors are pleased to acknowledge useful and
stimulating discussions with Profs. M. Braden, V. Kratvsov, 
Y. Maeno, A. Mackenzie, P. Johnson, and Yu Lu.



\begin{thebibliography}{40}
\bibitem{foot} The concept of a small polaron is here generalized slightly to
mean that the electron is dressed by a local cloud of neutral bosons, not
necessarily of phononic origin.
\bibitem{upturn} However, we should point out that these resistivity
measurements are carried out at constant pressure, while our
calculation is for constant volume. This upturn at high temperatures
may disappear once the thermal expansion is taken into account.
\end{thebibliography}
\end{document}